\documentclass{pasj02}
\usepackage[switch,mathlines]{lineno}
\usepackage{url}
\usepackage{graphicx}	
\usepackage{color}
\usepackage{mathrsfs}

\newcommand{\mbh}{M_{\rm BH}}
\newcommand{\msun}{M_\odot}

\newcommand{\mdot}{\dot m}

\newcommand{\mum}{\mu {\rm m}}

\newcommand{\Te}{\theta_{\rm e}}
\newcommand{\taue}{\tau_{\rm es}}

\newcommand{\K}{{\rm K}}
\newcommand{\beq}{\begin{equation}}
\newcommand{\eeq}{\end{equation}}

\Received{$\langle$reception date$\rangle$}
\Accepted{$\langle$acception date$\rangle$}
\Published{$\langle$publication date$\rangle$}

\begin{document}

\title{Weakness of X-rays and Variability in High-redshift AGNs with Super-Eddington Accretion}

\author{Kohei Inayoshi$^{1*}$}
\orcid{0000-0001-9840-4959}
\email{inayoshi@pku.edu.cn}

\author{Shigeo S. Kimura$^{2,3}$}
\orcid{0000-0003-2579-7266}

\author{Hirofumi Noda$^3$}
\orcid{0000-0001-6020-517X}

\altaffiltext{1}{Kavli Institute for Astronomy and Astrophysics, Peking University, Beijing 100871, China}
\altaffiltext{2}{Frontier Research Institute for Interdisciplinary Sciences, Tohoku University, Sendai 980-8578, Japan}
\altaffiltext{3}{Astronomical Institute, Graduate School of Science, Tohoku University, Sendai 980-8578, Japan}


\maketitle

\begin{abstract}
The James Webb Space Telescope (JWST) observations enable the exploration of active galactic nuclei (AGNs) 
with broad-line emission in the early universe. 
Despite their clear radiative and morphological signatures of AGNs in rest-frame optical bands, complementary evidence 
of AGN activity -- such as X-ray emission and UV/optical variability -- remains rarely detected. 
The weakness of X-rays and variability in these broad-line emitters challenges the conventional AGN paradigm, indicating that the accretion processes 
or environments around the central black holes (BHs) differ from those of low-redshift counterparts.
In this work, we study the radiation spectra of super-Eddington accretion disks enveloped by high-density coronae. 
Radiation-driven outflows from the disk transport mass to the poles, resulting in moderately optically-thick, warm coronae formed 
through effective inverse Comptonization.
This mechanism leads to softer X-ray spectra and larger bolometric correction factors for X-rays compared to typical AGNs, while being consistent
with those of JWST AGNs and low-redshift super-Eddington accreting AGNs.
In this scenario, UV/optical variability is suppressed due to photon trapping within super-Eddington disks, while X-ray emissions remain weak 
yet exhibit significant relative variability.
These characteristics are particularly evident in high-redshift AGNs powered by lower-mass BHs with $\lesssim 10^{7-8}~\msun$, which undergo
rapid mass accretion following overmassive evolutionary tracks relative to the BH-to-stellar mass correlation in the local universe.
\end{abstract}

%\pagewiselinenumbers

%%%%%%%%%%
%	Section 1     %
%%%%%%%%%%
\section{introduction}

Active galactic nuclei (AGNs) are key building blocks of massive black holes (BHs) in the present-day universe.
The James Webb Space Telescope (JWST) has revealed low-luminosity AGNs ($\sim 10^{44-46}~{\rm erg~s}^{-1}$) 
powered by low-mass BHs with $\mbh\sim 10^{6-8}~\msun$ at $z\sim 4-7$ (e.g., \cite{Onoue_2023,Kocevski_2023,Harikane_2023_agn,Maiolino_2023_JADES,Maiolino_2024a,Matthee_2024,Greene_2024}), 
providing constraints on their seeding and growth mechanisms at early cosmic time (e.g., \cite{Inayoshi_ARAA_2020,Volonteri_2021,Trinca_2022,W.Li_2024}).

JWST NIRSpec observations enable the measurement of broad H$\alpha$/H$\beta$ emission lines and confirm the presence of 
massive BHs in galactic nuclei. 
To date, most broad-line AGNs identified with JWST show extremely red colors in the rest-frame optical band 
\citep{Matthee_2024,Greene_2024,Kocevski_2024,Akins_2024}.
These AGNs, characterized by their red colors and compact morphology, are classified as ``Little Red Dots" (hereafter LRDs).
LRDs feature a distinctive v-shaped spectral energy distribution (SED) with red optical continua
presumably due to dust reddening and flux excesses at rest-frame UV bands \citep{Labbe_2023,Noboriguchi_2023,Barro_2024,Furtak_2024}.
However, MIRI observations have found faint rest-frame near-infrared emission at $\sim 1-2~\mum$
\citep{Williams_2023,Perez-Gonzalez_2024,Akins_2024}, where hot dust emission from AGN tori typically dominates with red continua.
The absence of strong hot dust emission implies that AGN contributions to LRD luminosities may be weak, though it could also indicate 
unrelaxed dust-flow structures during their early evolutionary stages \citep{Li_LRD_2024}.
Additionally, \citet{Kokubo_Harikane_2024} reported the lack of significant flux variability in sources exhibiting broad H$\alpha$ emission,
posing a challenge to the AGN interpretation for these newly-identified populations.
Recently, \citet{Zhang_2024} examined the variability significance in a sample of $\sim 300$ LRDs
and found that the LRD population on average does not show strong variability.
However, they identified eight strongly variable LRD candidates ($\simeq 2.7\%$ of the parent sample) with variability amplitudes of $0.24-0.82$ mag.

Over recent decades, X-ray observations have played a crucial role in studying AGNs embedded in dense circum-nuclear media 
(e.g., \cite{Ueda_2014,Aird_2015,Nandra_2015,Vito_2018,Ricci_2017,Ichikawa_2019,Ricci_2023}) due to the transmitted nature of high-energy radiation.
However, an intriguing aspect of JWST-identified AGNs is their non-detection of X-rays \citep{Kocevski_2023,Yue_2024,Juodzbalis_2024}. 
This holds true even with deep Chandra X-ray observations of individual sources and stacking analyses that push flux detection limits 
to $10^{-15}~{\rm erg~s}^{-1}~{\rm cm}^{-2}$ in the $2-10$ keV band \citep{Maiolino_2024b,Akins_2024}.
For these high-$z$ AGNs, the X-ray upper limits suggest $\alpha_{\rm ox}$ values (the flux ratios between UV at 2500~{\rm \AA} and hard X-rays
at 2 keV, defined by $\alpha_{\rm ox}\equiv {\rm d}\log F_\nu/{\rm d}\log \nu$) lower than those typical for AGNs ($\alpha_{\rm ox}\simeq -1.5$),
approaching the values seen in bright quasars ($\alpha_{\rm ox}<-1.8$) (e.g., \cite{Duras_2020}).
Remarkably, this X-ray weakness appears both in unobscured AGNs and LRDs independent of external obscuration levels
\citep{Maiolino_2024b}, indicating that this X-ray weakness could be intrinsic.
Despite the origin remaining unclear, the X-ray luminosity relative to the bolometric luminosity aligns with 
those of local super-Eddington accreting AGNs (e.g., \cite{Wang_2004,Laurenti_2022,Tortosa_2023}), some narrow-line 
Seyfert 1 galaxies (NLSy1; e.g., \cite{Liu_2021}), and nearby intermediate massive BHs (e.g., \cite{Dong_2012}).
Such X-ray weak AGNs have been previously identified, particularly among nearby sources with 
weak UV emission lines \citep{Wu_2012}.

These observations provide valuable insights into theoretical models of BH accretion and radiative processes in JWST AGNs.
Some studies have proposed that the X-ray weakness is explained by the geometrically-thick structures of super-Eddington accretion disks,
such as high inclination angles of observers from the poles \citep{Wang_2014,Pacucci_Narayan_2024} or warm coronae cooled by soft photons in a funnel-like 
reflective geometry \citep{Madau_Haardt_2024}.
Additionally, soft X-ray spectra associated with super-Eddington accretion may also explain the weakness of UV emission lines 
observed in some LRDs (\cite{Lambrides_2024}; see also \cite{Akins_2024b}).

An independent line of evidence supporting the idea that JWST-identified AGNs are powered by super-Eddington accreting BHs 
comes from the measurement of blue-shifted, prominent absorption on top of broad emission lines \citep{Matthee_2024,Maiolino_2023_JADES}.
Such striking spectral features are observed for at least $\sim 20\%$ of broad Balmer-line emitters including both LRDs and unobscured sources,
distinguishing these high-$z$ sources from their low-$z$ counterparts \citep{Maiolino_2023_JADES, Kocevski_2024,Lin_2024,Juodzbalis_2024}.
The detection of Balmer absorption indicates the presence of extremely high-density gas absorbers surrounding these AGNs with significant covering factor,
suggesting outflows carrying mass at super-Eddington rates \citep{Inayoshi_Maiolino_2024}.
These powerful, dense outflows are considered to be driven by outward radiation pressure from super-Eddington accreting disks,
as consistently found in radiation(-magneto) hydrodynamic simulations of BH accretion (e.g., \cite{Ohsuga_2005,Jiang_2014,Sadowski_2015,Hu_2022a}).

In this work, we propose an SED model to explain the X-ray weakness observed in JWST-selected AGNs by examining the structure of coronae 
surrounding super-Eddington accretion disks, which are accompanied by powerful outflows.
Effective mass loading in the corona regions enhances inverse Compton scattering efficiency and lowers the plasma temperature, 
resulting in a softer X-ray spectrum compared to typical AGNs.
Warm corona models have previously been explored in theoretical studies of certain AGNs and X-ray binary systems 
(e.g., \cite{Kubota_Done_2019,Kawanaka_Mineshige_2024}).
Our work extends this theoretical framework to high-redshift AGNs, incorporating modifications to suit this context.
Our SED model successfully accounts for the non-detection of X-rays in JWST-identified AGNs with lower-mass BHs, likely 
accreting at or above the Eddington rate.
We also explain the distribution of the X-ray bolometric correction factor across various types of objects at different redshifts and luminosities. 
Additionally, we propose that super-Eddington accretion can explain the absence of flux variability in the UV/optical bands due to photon trapping. 
This mechanism weakens the luminosity response to fluctuations in the mass accretion rate, while the X-ray flux variations 
show an anti-correlation with those in the UV/optical bands.

This paper is organized as follows. 
In Section~\ref{sec:compt}, we begin by reviewing the physics of Comptonization.
In Section~\ref{sec:XSED}, we construct broadband SEDs of super-Eddington accreting BHs (disk + warm coronae).
We then discuss the X-ray weakness of JWST-identified AGNs and the distribution of the hard X-ray bolometric correction 
(Section~\ref{sec:xray_weak}), the absence of significant flux variations (Section~\ref{sec:variability}), and the unique appearance of these characteristics preferentially at higher redshifts (Section~\ref{sec:unique}).
Section~\ref{sec:summay} summarizes this work.

%%%%%%%%%%
%	Section 2     %
%%%%%%%%%%
\vspace{-3mm}
\section{Comptonization in warm plasma}\label{sec:compt}

In this section, we introduce the physical processes underlying X-ray production due to Comptonization in hot/warm plasma \citep{Sunyaev_Titarchuk_1980,Haardt_Maraschi_1991,Haardt_Maraschi_1993,Titarchuk_1994}.
We do not provide detailed models of corona structures around accreting BHs (e.g., corona heating mechanisms and geometry), 
but instead focus on the robust relationship among physical quantities defined by the microphysics of Comptonization 
\citep{Rybicki_Lightman_1986}.

JWST-selected AGNs, powered by accreting massive BHs with masses of $\mbh\simeq 10^{6-8}~\msun$, typically have accretion-disk 
temperatures of $\gtrsim 10^{5-6}~\K$, making them bright in UV bands.
Therefore, emission with a harder spectrum toward X-ray regimes is produced from inverse Compton scattering by hot plasma surrounding the BH.
The spectral shape is determined by Comptonization of seed photons at a temperature $kT_{\rm seed}$ 
within hot electrons at temperature $kT_{\rm e}$ and optical depth $\taue =n_{\rm e}\sigma_{\rm es}r_{\rm cor}$, where $r_{\rm cor}$ 
is a characteristic scale  of the corona. 
A key quantity in this process is the Compton $y$-parameter defined as $y=(4\Te + 16\Te^2)\taue(\taue+1)$,
where $\Te=kT_{\rm e}/(m_{\rm e} c^2)$.
The dependence of the spectral index $\Gamma$, where $F_\nu \propto \nu ^{1-\Gamma}$, on the corona properties has been extensively studied. 
Here, we adopt a formula to quantify the photon index $\Gamma$ as a function of $\Te$ and $\taue$ 
(\cite{Titarchuk_Lyubarskij_1995}; hereafter TL95),
\begin{equation}
\Gamma \simeq -\frac{1}{2}+\sqrt{\frac{9}{4}+\frac{\beta}{\Te}},
\label{eq:Gamma1}
\end{equation}
where
\begin{equation}
\beta = \frac{\pi^2(1-e^{-0.7\taue})}{3(\taue+\frac{2}{3})^2}+e^{-1.4\taue}\ln \left(\frac{4}{3\taue}\right).
\label{eq:beta}
\end{equation}
This approximation applies to non-relativistic plasma with $\Te<1$ and assumes spherical geometry (see also \cite{Sunyaev_Titarchuk_1980}).
The form of $\beta$ is approximated so that it asymptotically approaches the optically thin and thick limits.
In the optically-thick limit ($\taue \gg 1$), Equations~(\ref{eq:Gamma1}) and (\ref{eq:beta}) reproduce the formula derived from
a self-similar solution of the steady-state Kompane’ets equation for non-relativistic regimes, $\Te\ll 1$ \citep{Shapiro_1976}.

For comparison, we introduce two additional formulae for calculating $\Gamma$ based on $\Te$ and $\taue$.
The first formula is given by \citet{Pozdniakov_1979} (hereafter P79),
\begin{align}
\Gamma \simeq &~ 1+ \frac{-\ln \taue +2/(3+\Te)}{\ln(12\Te^2+25\Te)},
\end{align}
which is applicable for $\taue \leq 3$ and $\Te>0.1$ \citep{Sunyaev_Titarchuk_1980}.
This form has been widely used in observation studies.
The second one is the simplest form proposed by \citet{Beloborodov_1999} (hereafter B99), 
\begin{equation}
\Gamma \simeq \frac{9}{4}y^{-2/9}.
\end{equation}
This expression was obtained for $1\lesssim y\lesssim 10$ by fitting the result of \citet{Coppi_1992} with 
two temperature cases of $\Te=0.1$ and $0.2$.

\begin{figure}
\begin{center}
\includegraphics[width=81mm]{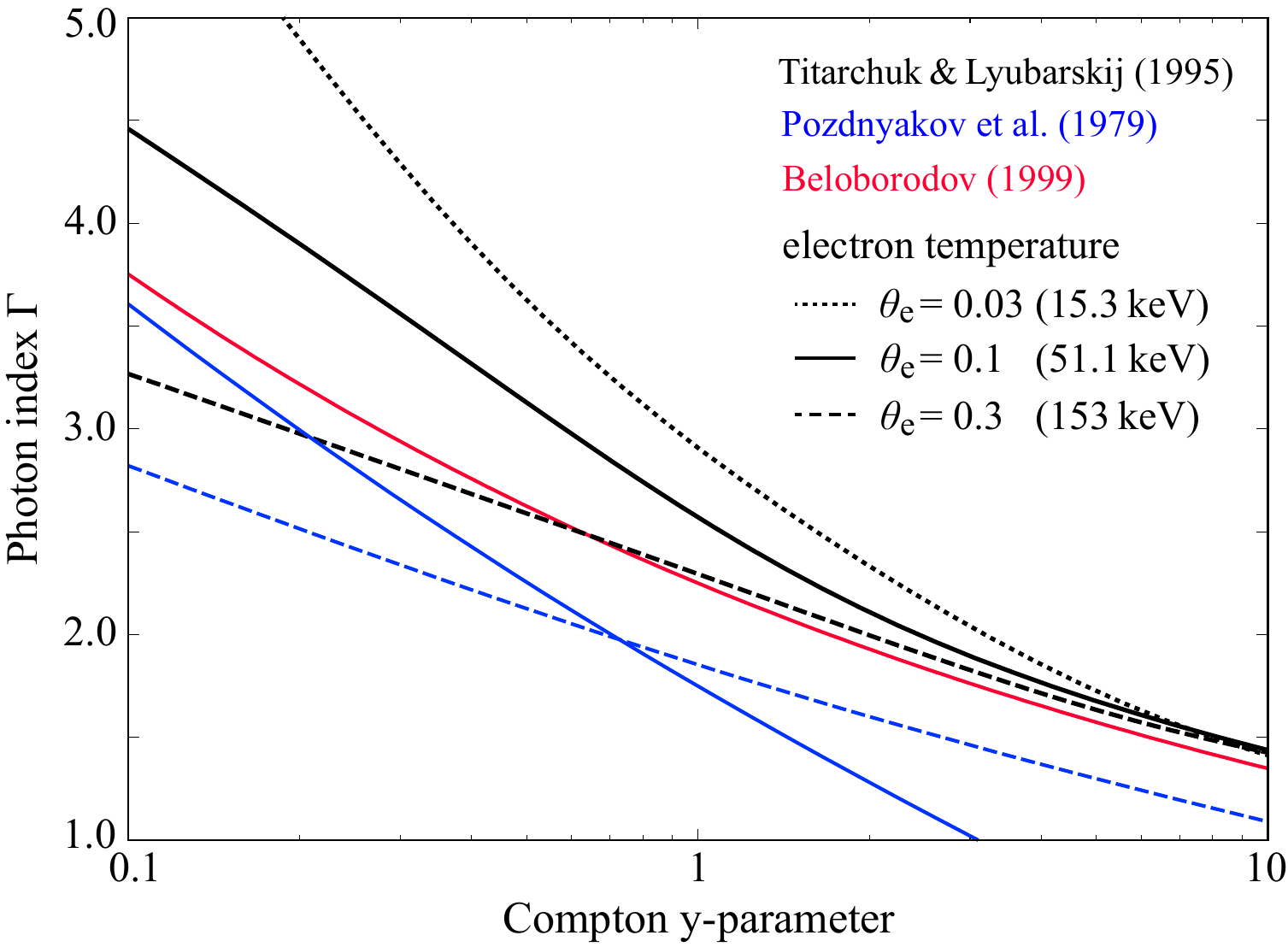}
\end{center}
\caption{X-ray spectral index ($\Gamma$; $F_\nu \propto \nu^{1-\Gamma}$) as a function of Compton $y$-parameter, based on three different studies;
\citet{Titarchuk_Lyubarskij_1995} (black, our fiducial model), \citet{Pozdniakov_1979} (blue), and \citet{Beloborodov_1999} (red). 
Three values of electron temperature in the non-relativistic regime are considered; $\Te=0.03$ (dotted), $0.1$ (solid), and $0.3$ (dashed).
Note that the fitting formula of \citet{Pozdniakov_1979} is not applied for $\Te\lesssim 0.1$.
{Alt text: Graphs showing the theory modeled curves from three different studies.}
}
\vspace{-5mm}
\label{fig:yGamma}
\end{figure}

Figure~\ref{fig:yGamma} shows the photon indices as a function of Compton $y$-parameter for the three different models: TL95 (black), P79 (blue), and B99 (red).
For the first two models, we consider various plasma temperatures of $\Te=0.03$ (dotted), $0.1$ (solid), and $0.3$ (dashed), where
the $y$-parameter varies with the optical depth $\taue$.
Despite the diversity of the fitting formulae and complex functional forms of the photon index, $\Gamma$ generally decreases (i.e., becomes harder) with increasing the $y$-parameter.
With the TL95 model, the photon index approaches $\Gamma \simeq 1$ in the limit of $y\gg 1$ (or $\taue \gg 1$ for a fixed $\Te$), 
while the index becomes as steep as $\Gamma \gtrsim 3$ due to inefficient Comptonization in the limit of $y\ll 1$.
The simplest form proposed by B99 shows good agreement with the TL95 model over a wide range of the $y$-parameter, except in the limit of $y\ll 1$.
In contrast, the P79 model shows discrepancies from the TL95 and B99 models for non-relativistic plasma ($0.1\leq \Te \leq 0.3$).
This may be because the P79 formula extends the form in relativistic regimes to moderate relativistic conditions, calibrated by Monte-Carlo radiation transfer.
Given our focus on non-relativistic plasma with $0.01\lesssim \Te \lesssim 0.3$ (see Section~\ref{sec:XSED}),
we adopt the TL95 model in this work.

%%%%%%%%%%
%	Section 3     %
%%%%%%%%%%
\vspace{-3mm}
\section{Spectra of super-Eddington accreting AGN from UV to X-rays}\label{sec:XSED}

The physics of Comptonization gives a relationship between the three key quantities of $\Te$, $\taue$, and $\Gamma$ (or the Compton $y$-parameter
given by $\Te$ and $\taue$).
This requires two more additional conditions to determine the X-ray spectral shape.

First, following \citet{Haardt_Maraschi_1991} and \citet{Haardt_Maraschi_1993} (see also \cite{Kawanaka_Mineshige_2024}), we consider the energy balance 
between the corona and disk region that provides soft seed photons for Comptonization.
Let us assume that a fraction $f_w$ of the accretion energy is dissipated within the corona and heats the region at a rate of $Q^+_{\rm cor}=f_w Q^+_{\rm tot}$ per unit surface area.
The corona region cools via Compton scattering of soft lower-energy photons $Q_{\rm cor}^-=yF_{\rm soft}$.
Thus, the energy equation ($Q^+_{\rm cor}=Q^-_{\rm cor}$) in the corona is given by 
\begin{equation}
    f_w Q^+_{\rm tot} = yF_{\rm soft},
    \label{eq:energy_corona}
\end{equation}
where we assume that the energy loss from the corona region is dominated through Compton cooling (see also \cite{Kawanaka_Mineshige_2024}).
On the other hand, the energy equation in the disk region is expressed as
\begin{equation}
    (1-f_w) Q^+_{\rm tot} + yF_{\rm soft}/2 = F_{\rm soft},
    \label{eq:energy_soft}
\end{equation}
where a half of the corona emission is used as heating of the disk surface\footnote{
A recent study by \citet{Madau_Haardt_2024} discussed the properties of coronae formed within a funnel-like region
surrounded by a geometrically thick disk at a super-Eddington accretion rate, where the density of soft-seed photons is 
enhanced within such a reflection geometry. In the Appendix, we provide a detailed comparison of this effect with our model.}.
We note that the equations governing the energetics between the disk and corona do not explicitly depend on the specific geometrical configuration of the system. 
However, they are indirectly influenced through the Compton $y$-parameter and the coronal heating efficiency $f_w$, which are determined by the assumed geometry.
With Equations~(\ref{eq:energy_corona}) and (\ref{eq:energy_soft}), we have a closure relation of
\begin{equation}
    y=\frac{2f_w}{2-f_w}.
    \label{eq:fwy}
\end{equation}
Given that $f_w\leq 1$, the Compton $y$-parameter is limited below $y\leq 2$.
In this formulation, the energy flux ratio of hard and soft photons is simply given by $y$.
The specific values of $y$ and $f_w$ can be determined once the physical mechanism is specified under an assumed geometry.
In this study, we adopt a Compton $y$-parameter on the order of unity, near its upper limit,
as motivated by radiation hydrodynamic simulations of Comptonization in radiation pressure-driven outflows from super-Eddington accretion disks
around BHs with $10\leq M_{\rm BH}/\msun\leq 10^4~\msun$ (e.g., \cite{Kawashima_2009, Kawashima_2012, Kitaki_2017}).
While in reality, the Compton $y$-parameter (or corona heating efficiency $f_w$) is not necessarily fixed in the system, we here explore the qualitative behavior 
of the X-ray SEDs of super-Eddington accreting supermassive BHs in AGNs within a certain range of $2/3 \leq y \leq 1$.

%%%%%%%%
%   Figure 1    %
%%%%%%%%

Second, we estimate the mass density and optical depth of the corona region, potentially embedded within outflows associated with nuclear accretion disks.
We assume that accretion flows involve co-existing inflows and outflows (including meridional circulation, e.g., convective motion; 
see \cite{Quataert_2000}), where the mass inflow rate measured at a spherical radius $r$ integrated over solid angles follows a power-law form, 
$\dot{M}_{\rm in}(r)\propto r^p$, with an index $0.5\lesssim p<1$ (e.g., \cite{Yuan_Narayan_2014,Inayoshi_2018,Hu_2022a,Xu_2023}).
Such outflows are commonly observed in numerical simulations of advection dominated accretion flows onto BHs,
both at sub- and super-Eddington rates, where advection cooling predominates over radiative cooling.
In a quasi-steady accretion flow, the net accretion rate $\dot{M}_{\rm BH}$ remains constant with radius and is normalized to the inflow rate 
at the outflow launching radius $r_0$.
Thus, the angle-integrated outflow rate can be expressed as $\dot{M}_{\rm out}(r)=\dot{M}_{\rm in}(r)-\dot{M}_{\rm BH}$, or
\begin{align}
    \dot{M}_{\rm out}=\dot{M}_{\rm BH} \left[\left(\frac{r}{r_0}\right)^p -1 \right],
\end{align}
which is valid at $r\geq r_0$.
Applying the continuity equation, the density of the outflow is given by
\begin{align}
    \rho_{\rm out}(r)=\frac{\dot{M}_{\rm out}}{\Omega v_{\rm esc}r^2}=\frac{\dot{M}_{\rm BH}}{\Omega c r_{\rm sch}^2}f_p(x),
\end{align}
where $r_{\rm sch}=2GM_{\rm BH}/c^2$ is the Schwarzschild radius, $v_{\rm esc}=\sqrt{2G\mbh/(r-r_{\rm sch})}$ is the escape velocity from a radius of $r$, 
$\Omega$ is the solid angle within which outflows are injected, 
$f_p(x)=[(x/x_0)^p-1]\sqrt{x-1}/x^2$, and $x=r/r_{\rm sch}$ ($x_0=r_0/r_{\rm sch}$).
Given the anisotropic level of the outflow component, one can calculate the outflow-loading solid angle $\Omega$ \citep{Takeo_2020}. 
When the angular distribution of the outflow rate follows a dependence of $\propto (\cos \theta)^N$, the density along the polar direction is 
enhanced by a factor of $(N+1)$ due to the effect of collimation. In this work, we adopt a moderate level of anisotropy with $N=1$, 
corresponding to a solid angle of $\Omega = 2\pi$, and the optical depth is calculated as 
\begin{align}
    \taue=\int_{\rm r_0}^{r_{\rm cor}} \rho_{\rm out}\kappa_{\rm es}dr = \frac{10\dot{m}_{\rm BH}}{(\Omega/2\pi)}
    \int_{x_0}^{x_{\rm cor}} f_p(x)dx,
    \label{eq:tau}
\end{align}
where $\dot{m}_{{\rm BH}}(\equiv \dot{M}_{\rm BH}/\dot{M}_{\rm Edd})$ is the BH accretion rate normalized by the Eddington rate defined by
$\dot{M}_{\rm Edd}\equiv L_{\rm Edd}/(0.1c^2)$ and $L_{\rm Edd}$ is the Eddington luminosity (see the notation in e.g., \cite{Inayoshi_ARAA_2020}).
While the outflow launching radius varies across studies, it is generally located just outside the innermost stable circular orbit (ISCO) for a given BH spin
\citep{Sadowski_2015}. For a non-rotating BH, $r_0\simeq r_{\rm ISCO}=3~r_{\rm sch}$ ($x_0=3$).

For super-Eddington accretion disks, the power-law index is quantified as $p\sim 0.5-0.7$ by radiation hydrodynamic simulations for long-term evolution 
of BH accretion \citep{Hu_2022a}, with weak dependence on the outer boundary and initial conditions of the simulation. 
Thus, the integral part $\mathscr{F}_p$ is approximated as 
$\mathscr{F}_p\simeq 0.16-0.37$ for $p=0.5$ and $\simeq 0.24-0.59$ for $p=0.7$, when the corona size ranges over $10\leq r_{\rm cor}/r_{\rm sch}\leq 20$. 
Assuming a corona size of $\sim 10~r_{\rm sch}$ and an outflow covering half the sky, the optical depth simplifies to 
\begin{equation}
\taue\simeq 2\dot{m}_{\rm BH}\left(\frac{\mathscr{F}_p}{0.2}\right)\left(\frac{\Omega}{2\pi}\right)^{-1}.
\label{eq:tau1}
\end{equation}
The optical depth scales with the Eddington ratio of the BH accretion rate.
While the numerical factor depends on several variables such as $p$, $r_0$, $r_{\rm cor}$, and $\Omega$, the conclusion that super-Eddington flows create 
moderate optically-thick regions to the poles with a quasi-spherical geometry remains valid \citep{Chang_Ostriker_1985}.

A caveat remains regarding the efficiency of mass loading into coronae from radiatively efficient accretion disks at moderately 
high accretion rates $\dot{M}_{\rm in}/\dot{M}_{\rm Edd}\sim O(10^{-1})$, where radiative cooling balances with viscous heating
\citep{SS_1973}. 
Numerical simulations predict that geometrically-thin disks produce weaker outflows in this regime ($p\simeq 0.1-0.2$; \cite{Jiang_2019,Inayoshi_2019}).
For these low values of $p$, one obtains $\mathscr{F}_p\simeq (3-6)\times 10^{-2}$ for $x_{\rm cor}=10$, which is $\simeq 4-5$
times lower than in our fiducial super-Eddington case.
Moreover, in this accretion-rate regime, the hard X-ray emission observed in some sources can be attributed to
hot accretion flows in the inner region of a truncated accretion disk or to slab-like hot plasma layers 
above a geometrically-thin cold disk \citep{Meyer_Meyer-Hofmeister_1994,Liu_1999} rather than the outflows 
discussed in this work (see Section~\ref{sec:XSED} for details).
In contrast, observations of X-ray binaries suggest that substantial mass is expelled from accretion disks regardless of their thermal state 
(e.g., \cite{Tetarenko_2018}), potentially driven by magnetically-driven outflows even in radiatively cooled disks 
(e.g., \cite{Suzuki_Inutsuka_2009,Ohsuga_2011,Bai_2013,Dihingia_2023}).
To account for these uncertainties, our model assumes a constant outflow strength with $p=0.5$ ($\mathscr{F}_p=0.2$) independent of accretion rate, 
though lower values of $p$ (or $\mathscr{F}_p$) can be assigned to represent moderately sub-Eddington rates when specified.

%%%%%%%%
%   Figure 2    %
%%%%%%%%
\begin{figure}
\begin{center}
\includegraphics[width=83mm]{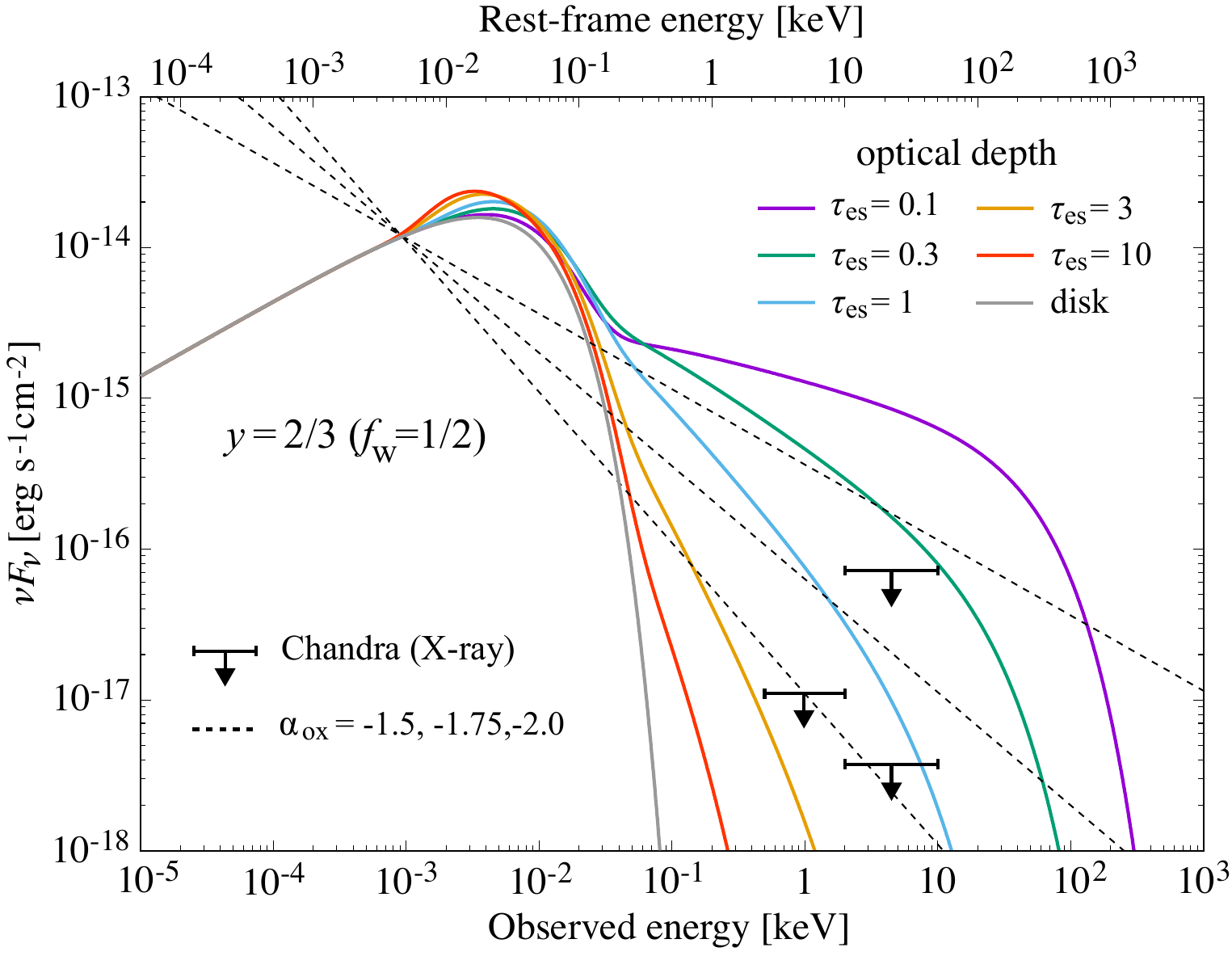}
\end{center}
\caption{Broadband SEDs of a $z=4$ AGN  with a bolometric luminosity of $L_{\rm bol}=10^{46}~{\rm erg~s}^{-1}$ for different values of the optical depths ($0.1\leq \taue \leq 10$). 
The Compton parameter is set to $y=2/3$. The gray curve shows the SED of disk seed photons.
For reference, we overlay the upper bound of the stacked AGNs obtained by Chandra observations (black arrows; \cite{Maiolino_2024b, Akins_2024}), and
single power-law spectra with indices of $\alpha_{\rm ox}=-1.5$, $-1.75$, and $-2.0$ (dashed lines).
{Alt text: Graphs showing the results of our SED model and the observational data.}}
\label{fig:SED_y2/3}
\vspace{-3mm}
\end{figure}

In this work, we consider the spectral shape of an AGN with two components: (1) an accretion disk with a big blue bump temperature $T_{\rm b}$
and (2) non-thermal corona emission characterized by a photon index $\Gamma$ and cutoff energy $kT_{\rm e}$.
The two components are calculated as 
\begin{equation}
F_{{\rm UV}}(\nu) \propto \nu^{\alpha_{\rm UV}}\exp \left(-\frac{h\nu}{k T_{\rm b}}\right),
\end{equation}
and
\begin{equation}
F_{{\rm X}}(\nu) \propto  \nu^{1-\Gamma}\exp \left(-\frac{h\nu}{kT_{\rm e}}-\frac{kT_{\rm b}}{h\nu}\right)
\end{equation}
where we set the UV spectral index to $\alpha_{\rm UV}=-0.5$, consistent with that of the low-redshift composite quasar SED \citep{VandenBerk_2001}. 
The radial profile of the disk surface temperature along the equator ($\theta=\pi/2$) is calculated with an analytical expression for Slim disk solutions 
described in \citet{Watarai_2006},
\begin{equation}
T_{\rm eff}(R) = 7.91\times 10^6 f^{1/8} M_{{\rm BH},7}^{-1/4} \hat{R}^{-1/2}\mathcal{F}(R,\dot{m}_{\rm BH}),
\label{eq:temp}
\end{equation}
where $R$ is the cylindrical radius, $\hat{R}\equiv R/r_{\rm sch}$, $M_{{\rm BH},7}=\mbh/(10^7~\msun)$, 
$f$ is the ratio of the advection cooling rate to the viscous heating rate\footnote{The fitting form of the ratio is given by \citet{Watarai_2006} as 
$f=0.5(X^2+2-X \sqrt{X^2+2})$, where $X=0.281R/(\dot{m}_{\rm BH} r_{\rm sch})$.}, and the function $\mathcal{F}$ characterizes the disk-inner boundary as 
$\mathcal{F}(R,\dot{m}_{\rm BH}) = [1-\sqrt{r_{\rm in}/R}]^{1/4}$ for $\dot{m}_{\rm BH}<4$, and $\mathcal{F}(R,\dot{m}_{\rm BH})=1$ for $\dot{m}_{\rm BH}\geq4$.
The radius of the disk inner edge ($r_{\rm in}$) is set to the ISCO radius for a non-spinning BH ($r_{\rm in}=3~r_{\rm sch}$) 
for $\dot{m}_{\rm BH} \leq 1$, and for $1<\dot{m}_{\rm BH} <4$ is calculated with a linear interpolation in the plane of $\log \dot{m}_{\rm BH} - \log r_{\rm in}$ 
between the ISCO radius and $1.1~r_{\rm sch}$.
We consider the extension of the disk inner radius at high accretion-rate cases ($\dot{m}_{\rm BH} >1$) because the gas is optically thick even inside the ISCO
(see \cite{Watarai_2000, Watarai_2006, Abramowicz_2010}, also \cite{Lasota_2024}).
The big blue bump temperature $T_{\rm b}$ is set to the local maximum value of the disk temperature profile.
Note that the surface temperature profile reproduces the conventional form of $T_{\rm eff}(R)\propto R^{-3/4}$ for the sub-Eddington regime \citep{SS_1973}.
We also introduce a low-energy cutoff for the non-thermal spectrum in the form of $e^{-kT_{\rm b}/h\nu}$ to 
ensure that the frequency-integrated flux is finite.

%%%%%%%%
%   Figure 3    %
%%%%%%%%
\begin{figure*}
\begin{center}
\includegraphics[width=85mm]{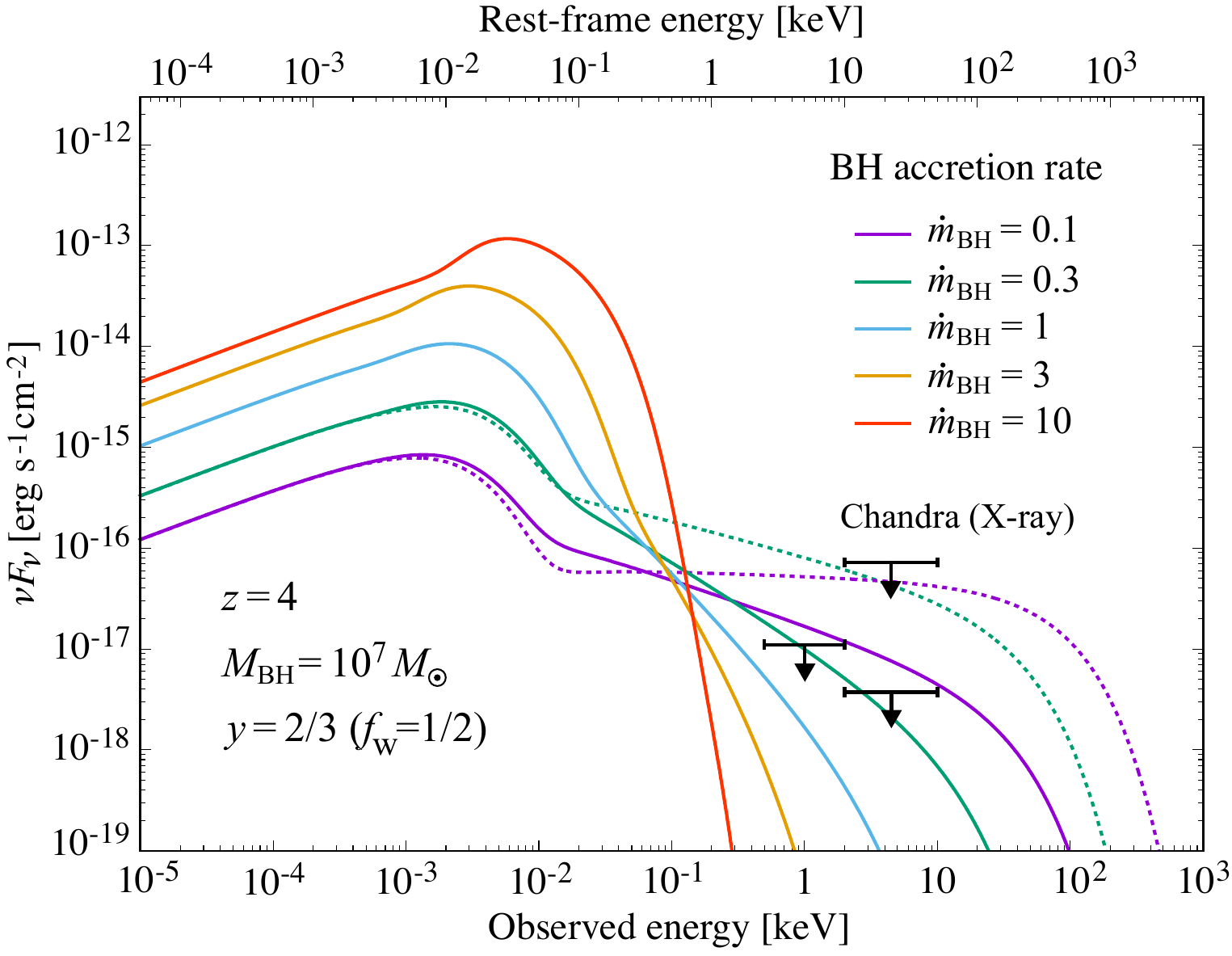}\hspace{3mm}
\includegraphics[width=85mm]{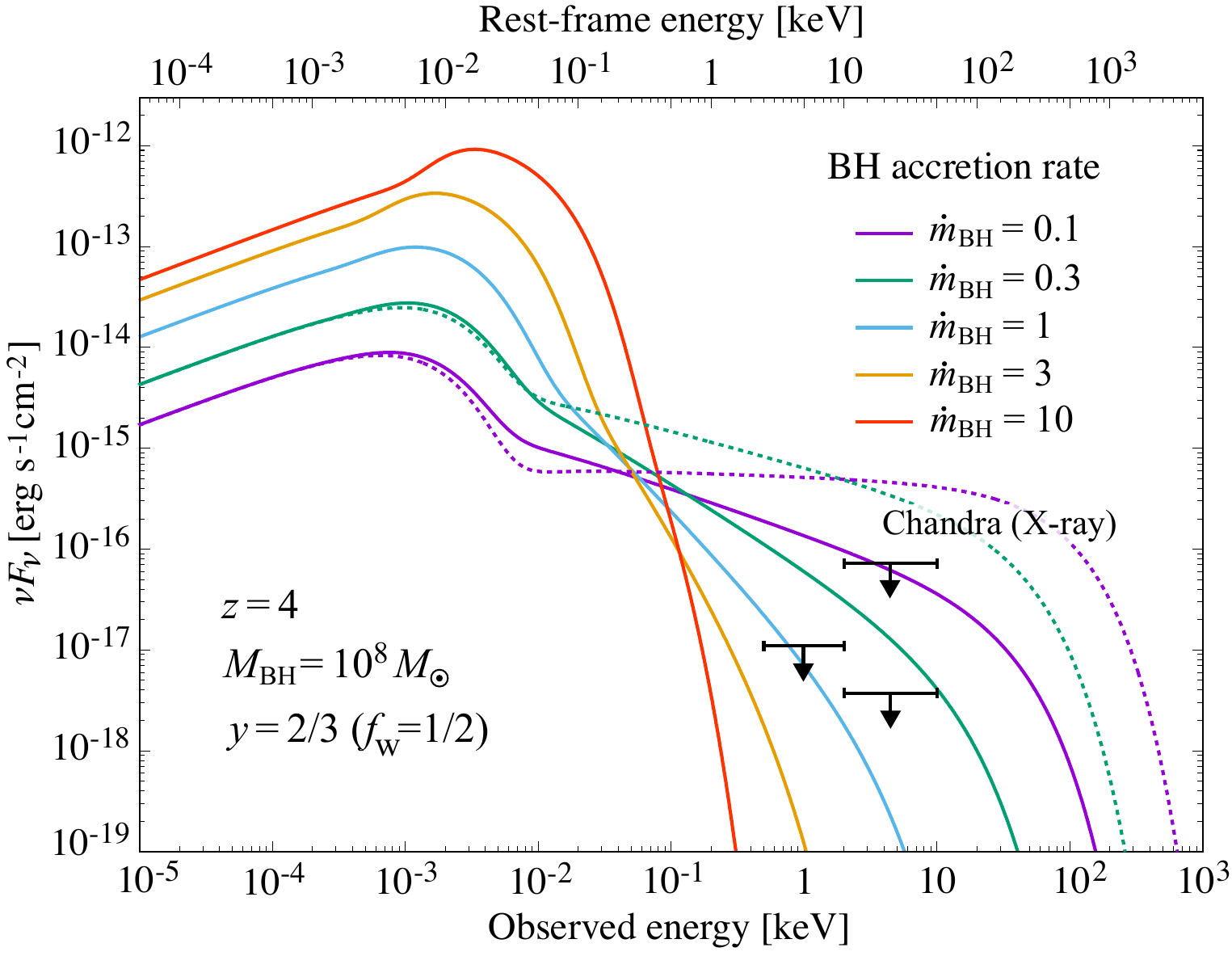}
\end{center}
\caption{Broadband SEDs of a $z=4$ AGN powered by a BH with $\mbh=10^7~\msun$ (left) and $10^8~\msun$ (right) accreting at 
various rates of $0.1\leq \dot{m}_{\rm BH} \leq 10$. Here, $L_{\rm bol}$, $T_{\rm b}$, $T_{\rm seed}$ are calculated consistently 
for given $\mbh$ and $\dot{m}_{\rm BH}$ values.
The Compton parameter is set to $y=2/3$, and optical depth is calculated by using Equation~(\ref{eq:tau}) with 
$\mathscr{F}_p=0.2$ (solid curves).
For sub-Eddington cases ($\dot{m}_{\rm BH}=0.1$ and $0.3$), SEDs with a lower mass-loading factor of 
$\mathscr{F}_p=0.05$ are presented (dotted curves), corresponding to lower-density coronae consistent with 
low-redshift AGNs that exhibit X-ray spectral hardness.
While the bolometric/UV luminosity increases with $\dot{m}_{\rm BH}$, the X-ray emission is sufficiently weak for $\dot{m}_{\rm BH} \gtrsim 1$ ($\mbh=10^7~\msun$) and 
$\dot{m}_{\rm BH} \gtrsim 3$ ($\mbh=10^8~\msun$), respectively, because of lower electron temperatures (and softer photon index) of the optically-thick corona. 
{Alt text: Graphs showing the results of our SED model and the observational data.}}
\label{fig:SED_M}
\vspace{-3mm}
\end{figure*}

We calculate the AGN disk luminosity by using the analytical formula \citep{Watarai_2000}\footnote{
This relationship between the accretion rate and disk luminosity deviates slightly from the formula in \citet{Watarai_2006}, 
which provide the basis for deriving the temperature profile we adopt. The largest deviation ($\sim 35\%$) occurs at $\dot{m}\simeq 2$. 
However, the discrepancy is negligible in the low- and high-accretion-rate regimes, and thus does not affect the main conclusions of our study.}:
\begin{equation}
\frac{L_{\rm disk}}{L_{\rm Edd}}=
\left\{
\begin{array}{ll}
\dot{m}_{\rm BH} & ~~~(\dot{m}_{\rm BH}<2), \\[6pt]
2\left[ 1+\ln \left(\dfrac{\dot{m}_{\rm BH}}{2}\right)\right] & ~~~(\dot{m}_{\rm BH}\geq 2). \\
\end{array}
\right.
\label{eq:Lbol}
\end{equation}
For simplicity, we set the monochromatic luminosity at 3000~\AA, representing the disk luminosity $L_{\rm disk}$, 
and express the bolometric luminosity as $L_{\rm bol} = 5.15~ \lambda L_{3000}$ \citep{Richards_2006}.
The normalization of the non-thermal component is adjusted so that
\begin{equation}
    \int_0^\infty F_{\rm X}(\nu) d\nu = y \int_{\nu_{\rm min}}^\infty F_{\rm UV}(\nu) d\nu,
\end{equation}
where the minimum energy of seed photons is set to $h\nu_{\rm min}=kT_{\rm seed}$ and $T_{\rm seed}$ is the disk temperature at 
$R=10~r_{\rm sch}~(\simeq r_{\rm cor})$.
In a Slim disk model, the disk-surface temperature structure follows $T_{\rm eff}\propto R^{-1/2}$ within the photon-trapping radius, and 
an optically-thick structure extends even into the interior of the ISCO down to the BH horizon at $\mdot_{\rm BH}\gg 1$.

In Figure~\ref{fig:SED_y2/3}, we show the SED of a $z=4$ AGN for optical depths ranging from $0.1$ to $10$. 
The Compton parameter is set to $y=2/3$, corresponding to a heating efficiency of the corona, $f_w=1/2$.
The gray curve represents the disk SED, which provides seed photons for Comptonization.
To demonstrate the SED dependence on the optical depth (excluding the $\dot{m}_{\rm BH}$ dependence),
we here set the bolometric luminosity to $L_{\rm bol}=10^{46}~{\rm erg~s}^{-1}$, the disk temperatures $T_{\rm b}=4\times 10^5~\K$, 
and the minimum temperature of seed photons for Comptonization $T_{\rm seed}=T_{\rm b}/2$ instead of 
using Equations~(\ref{eq:temp}) and (\ref{eq:Lbol}).
For comparison, we include the upper bound of the stacked AGNs obtained from Chandra X-ray observations \citep{Maiolino_2024b, Akins_2024}.
The dashed curves show single power-law spectra normalized at rest-frame $2500~{\rm \AA}$ with indices of $\alpha_{\rm ox}=-1.5$, $-1.75$, and $-2.0$ ($F_\nu \propto \nu ^{\alpha_{\rm ox}}$), connecting the UV and X-ray flux densities.

At the lowest optical depth ($\taue=0.1$), the photon index is approximately $\Gamma \simeq 2.24$, and the electron temperature reaches 
$\Te \simeq 0.5~(kT_{\rm e}\simeq 260~{\rm keV})$.
The X-ray SED shows a hard spectrum extending to the cutoff energy, with X-ray fluxes in the observed $2-10$ keV band reaching 
$\simeq 6\times 10^{-16}-10^{-15}~{\rm erg~s}^{-1}~{\rm cm}^{-2}$. 
As the optical depth increases (up to unity), the spectral shape becomes significantly softer due to a steeper photon index ($\Gamma \simeq 3$) 
and lower electron temperature ($kT_{\rm e} \simeq 34~{\rm keV}$) for $\taue=1$.
In these optically-thin regimes ($\taue\leq 1$), the X-ray flux normalization is consistent with values derived from assuming 
$-1.75 \lesssim \alpha_{\rm ox}\lesssim -1.5$.
Despite their softer X-ray spectra, the X-ray flux levels are comparable to those achievable by Chandra observations.

%%%%%%%%
%   Figure 4    %
%%%%%%%%
\begin{figure*}
\begin{center}
\includegraphics[width=125mm]{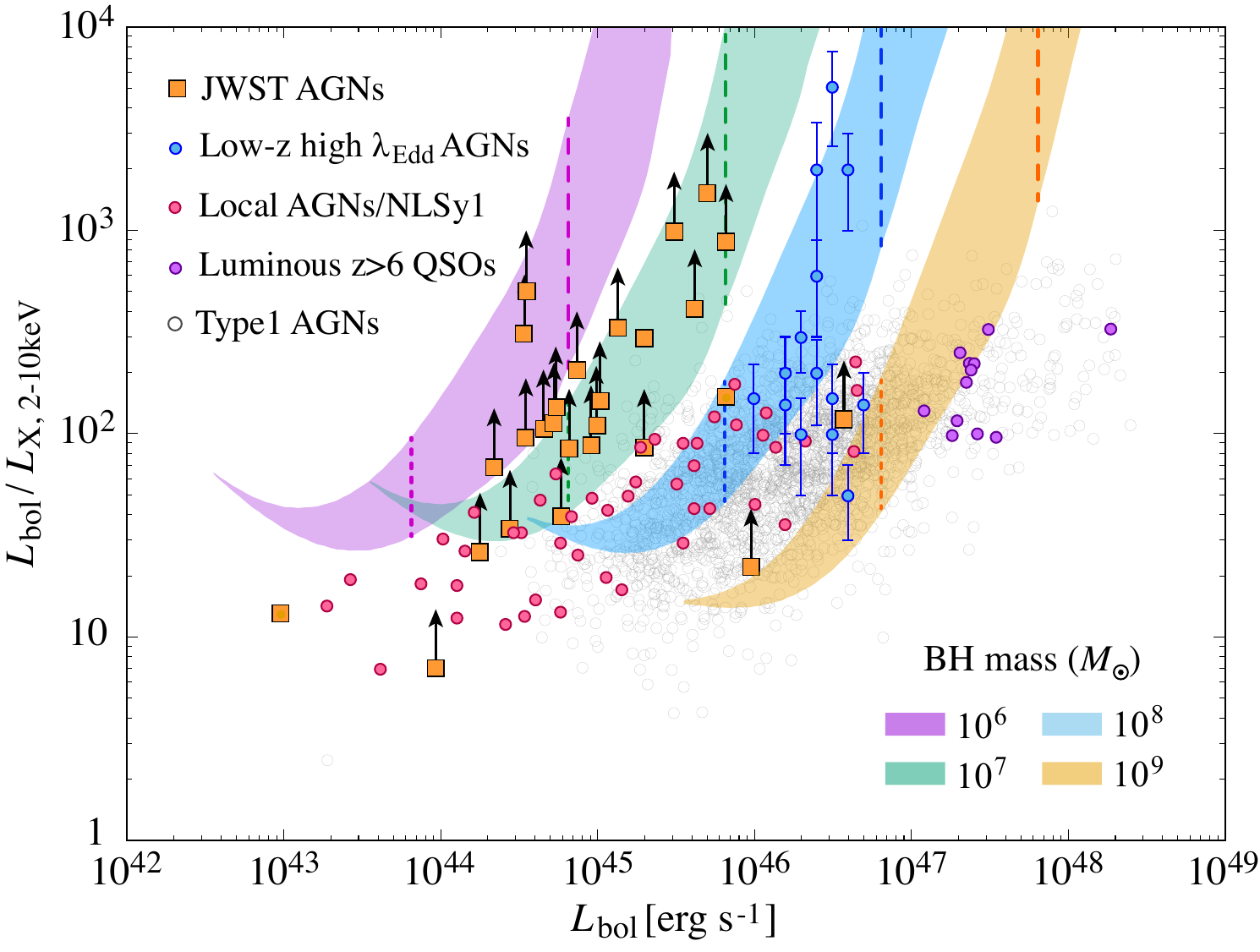}
\end{center}
    \caption{X-ray bolometric correction in the 2-10 keV band as a function of AGN bolometric luminosity for various BH masses 
    ($10^6 \leq \mbh/\msun \leq  10^{9}$) and accretion rates ($0.01 \leq \dot{m}_{\rm BH} \leq 10$).
    The short and long-dashed lines show the bolometric luminosities for $\dot{m}_{\rm BH} =0.1$ and $1.0$.
    For each BH mass, the Compton parameter is set at $2/3\leq y\leq 1$ (shaded region), where a larger $y$-value results in the higher X-ray luminosity 
    for the same bolometric luminosity.
    Observation data for various types of AGNs from literatures are overlaid for comparison: JWST-identified broad-line AGNs 
    (orange square; \cite{Maiolino_2024b}), low-redshift  high-$\lambda_{\rm Edd}$ quasars (blue circle; \cite{Laurenti_2022}), local AGNs including
    NLSy1 galaxies (red circle; \cite{Liu_2021}), luminous quasars at $z>6$ (purple circle; \cite{Zappacosta_2023}), and 
    optical/UV-selected typical quasars at $0<z<7$ (gray circle; \cite{Lusso_2020}).
    {Alt text: Graphs showing our modeled results and the observational data from the literature.}}
    \label{fig:kbol_x}
\vspace{-3mm}
\end{figure*}

In the optically-thick regime ($\taue>1$), where the electron temperature decreases as $\Te \propto \taue^{-2}$ (for a fixed $y$), the Comptonization 
spectrum does not extend well above $\sim 1~{\rm keV}$.
Finally, the X-ray spectrum becomes sufficiently soft and undetectable with the deep Chandra survey \citep{Yue_2024,Maiolino_2024b, Akins_2024}.
The low electron temperature in the optically-thick corona is consistent with findings in \citet{Kawanaka_Mineshige_2024}.

Figure~\ref{fig:SED_M} presents the SED of a $z=4$ AGN powered by a massive BH with $\mbh=10^7~\msun$ (left) and $10^8~\msun$ (right) accreting at 
various rates of $0.1\leq \dot{m}_{\rm BH} \leq 10$.
In these plots, $L_{\rm bol}$, $T_{\rm b}$, $T_{\rm seed}$ are calculated by using Equations~(\ref{eq:temp}) and (\ref{eq:Lbol})
for given $\mbh$ and $\dot{m}_{\rm BH}$ values.
The Compton parameter is set to $y=2/3$, and optical depth is calculated by using Equation~(\ref{eq:tau1}) with 
$\mathscr{F}_p =0.2$ and $\Omega=2\pi$ (solid curves).
Generally, the UV luminosity increases with the accretion rate, following $L_{\rm disk} \propto \dot{m}_{\rm BH}$ in the sub-Eddington and $\propto \ln \dot{m}_{\rm BH}$ 
in the super-Eddington regime, while the X-ray flux at the rest-frame $2-10$ keV band decreases with increasing the accretion rate.
The X-ray weakness results from a higher optical depth in the corona, which lowers the electron temperature and softens the Comptonization spectrum 
(see also Figure~\ref{fig:SED_y2/3}).
Although the bolometric luminosity increases with higher $\dot{m}_{\rm BH}$, the X-ray emission remains undetectable for $\dot{m}_{\rm BH} \gtrsim 1$ ($\mbh=10^7~\msun$) and $\dot{m}_{\rm BH} \gtrsim 3$ ($\mbh=10^8~\msun$), respectively.
This trend holds even with a larger Compton $y$-parameter, though a higher accretion rate is required to suppress the X-ray emission.

It is worth emphasizing that the UV spectrum becomes harder as the accretion rate increases, especially above the Eddington rate.
This occurs because the inner edge of the optically-thick accretion disk is not truncated at the ISCO; instead, it penetrates toward the BH event horizon 
when the accretion rate is sufficiently high (e.g., \cite{Watarai_2000, Abramowicz_2010,Kubota_Done_2019}). 
The harder spectrum arises from the temperature structure in a steady state of super-Eddington accretion disks, rather than being an artifact of the model.

At sub-Eddington accretion rates ($\dot{m}_{\rm BH}=0.1$ and $0.3$), we present the SEDs for a lower mass-loading factor of 
$\mathscr{F}_p=0.05$ (dotted curves), corresponding to lower-density coronae.
In this regime, reduced optical depths at a given accretion rate lead to higher electron temperatures and thus harder X-ray spectra
with photon indices of $\Gamma\simeq 2$, consistent with observations of X-ray AGNs at $0.01<z<0.5$ \citep{Trakhtenbrot_2017}. 
Alternatively, a larger Compton $y$-parameter ($y=1$) also produces similarly hard spectra.
This hard X-ray emission can be attributed to hot accretion flows in the inner region of a truncated accretion disk
or to slab-like hot plasma layers above a geometrically-thin cold disk (e.g., \cite{Meyer_Meyer-Hofmeister_1994,Liu_1999,Meyer_2000})
rather than to dense outflows we discuss in this work.
Multi-component corona models with hot and warm temperatures effectively reproduce the observed SEDs of 
moderately sub-Eddington sources (e.g., \cite{Jin_2012,Kubota_Done_2018}).
Additionally, some X-ray selected NLSy1 galaxies harboring super-Eddington accreting BHs show 
moderately hard X-ray components in their SEDs, possibly indicating the presence of hot coronae with 
lower mass-loading factors even in super-Eddington cases with $L_{\rm bol}/L_{\rm Edd} \sim 2$ (e.g., \cite{Jin_2012}).
Assuming a constant mass-loading factor of $\mathscr{F}_p=0.05$ for all accretion rates in our SED model, 
accretion rates above $\dot{m}_{\rm BH}\gtrsim 10(\mathscr{F}_p/0.05)^{-1}(\Omega/2\pi)$ are required to remain consistent 
with the Chandra upper limit
(see Equation~\ref{eq:tau1} and the critical optical depth of $\taue \gtrsim 2-3$ shown in Figure~\ref{fig:SED_y2/3}).
The lower mass-loading factors aligns well with the $L_{\rm bol}/L_{\rm Edd}-\Gamma$ relationship observed in local 
NLSy1 galaxies \citep{Liu_2021}, though the fiducial choice of $\mathscr{F}_p=0.2$ reproduces the same relationship broadly (see also Appendix~\ref{sec:LG}).
We also note that collimation of outflows is more likely at higher accretion rates ($\Omega<2\pi$ or $N\gtrsim 1$), 
particularly in the super-Eddington regime (e.g., \cite{Ohsuga_2005}).
Despite these caveats, we adopt a fixed mass loading factor of $\mathscr{F}_p=0.2$ 
independent of accretion rate, for simplicity in the following analysis.

%%%%%%%%%%
%	Section 4     %
%%%%%%%%%%
\vspace{-3mm}
\section{Discussion}

\subsection{X-ray weakness of rapidly accreting, massive BHs}
\label{sec:xray_weak}

In this section, we use the SED model developed in Section~\ref{sec:XSED} to demonstrate the X-ray weakness of AGNs identified through JWST observations.
Figure~\ref{fig:kbol_x} presents the ratio of bolometric luminosity ($L_{\rm bol}$) to $2-10$ keV X-ray luminosity ($L_{\rm X,2-10~keV}$) for various BH 
masses and accretion rates.
For each BH mass, we adopt Compton parameters at $2/3\leq y\leq 1$ (shaded regions), where a larger $y$-value results in higher X-ray luminosity 
for the same bolometric luminosity (i.e., a lower ratio on the vertical axis).
For comparison, we include data for various types of AGNs from the literatures: JWST-identified broad-line AGNs \citep{Maiolino_2024b}, 
low-redshift  high-$\lambda_{\rm Edd}$ quasars \citep{Laurenti_2022}, local AGNs including NLSy1 galaxies 
\citep{Liu_2021}, luminous quasars at $z>6$ \citep{Zappacosta_2023}, and optical/UV-selected typical quasars at $0<z<7$ \citep{Lusso_2020}.
We note that the bolometric correction factors from UV/optical bands used in the literature are not consistent.
Thus, when the UV luminosity or flux (e.g., $2500$ or $3000$~\AA) is available in these studies, 
we recalculate the bolometric luminosity
using the same bolometric correction factor adopted in this paper to ensure a fair comparison.

For BHs heavier than $10^8~\msun$, bolometric luminosities reach $L_{\rm bol}\simeq 10^{46-47}~{\rm erg~s}^{-1}$, consistent with 
bright quasar populations, assuming a typical Eddington ratio of $\dot{m}_{\rm BH} \simeq 0.3$ (e.g., \cite{Kollmeier_2006}).
The hard X-ray bolometric correction factors of $L_{\rm bol}/L_{\rm X,2-10~keV}\gtrsim 10-100$ align with those observed in pre-JWST AGNs.
In particular, our high-accretion models with $\mbh = 10^8~\msun$ agree with low-$z$, high-$\lambda_{\rm Edd}$ AGNs (blue circles; \cite{Laurenti_2022}),
where X-ray bolometric corrections reach as high as $\sim 10^3$ and BH masses range over $10^{7.9}\leq \log(\mbh/\msun) \leq 10^{8.5}$.

%%%%%%%%
%   Figure 5    %
%%%%%%%%
\begin{figure*}
\begin{center}
\includegraphics[width=83mm]{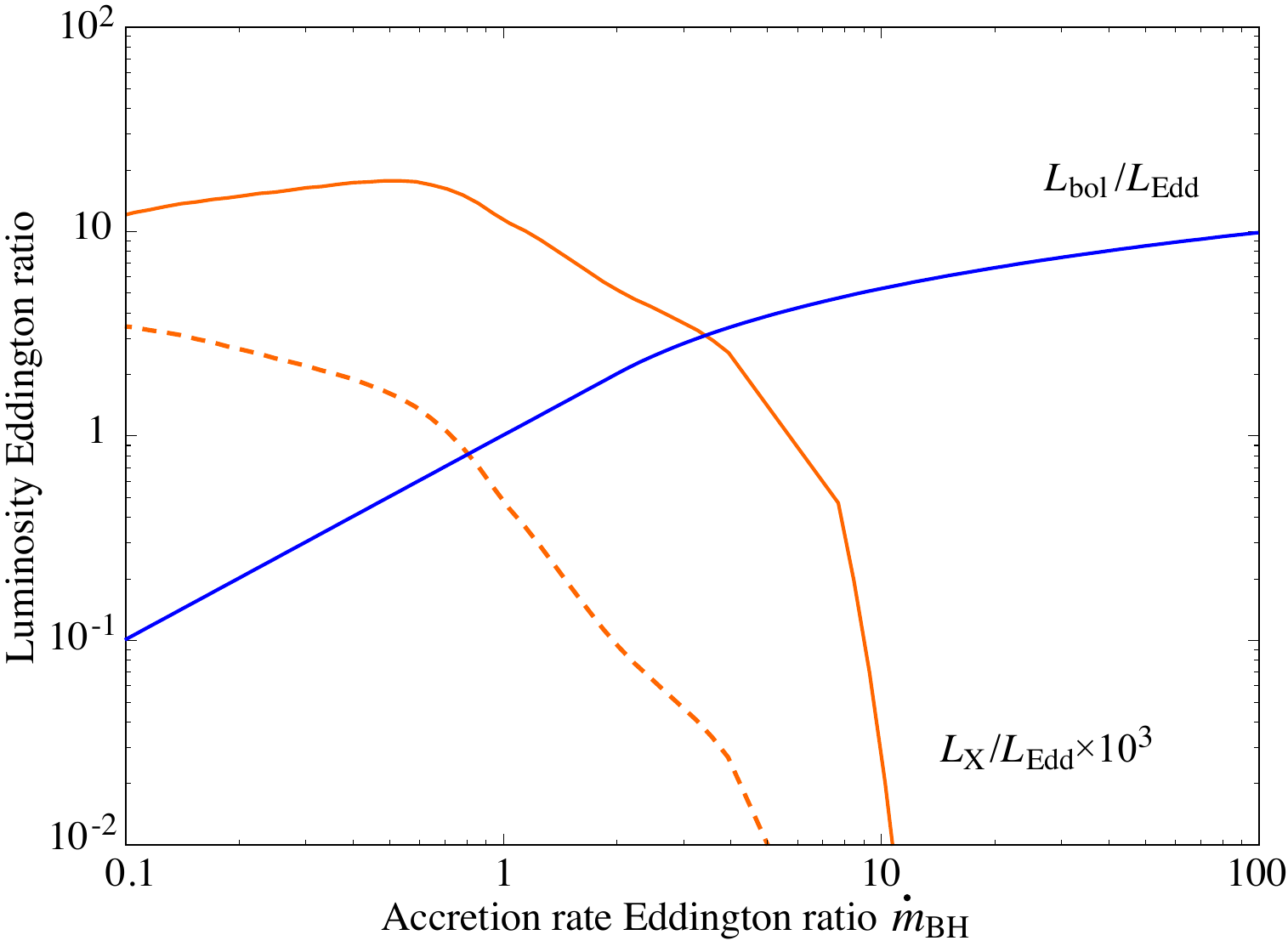}\hspace{3mm}
\includegraphics[width=83mm]{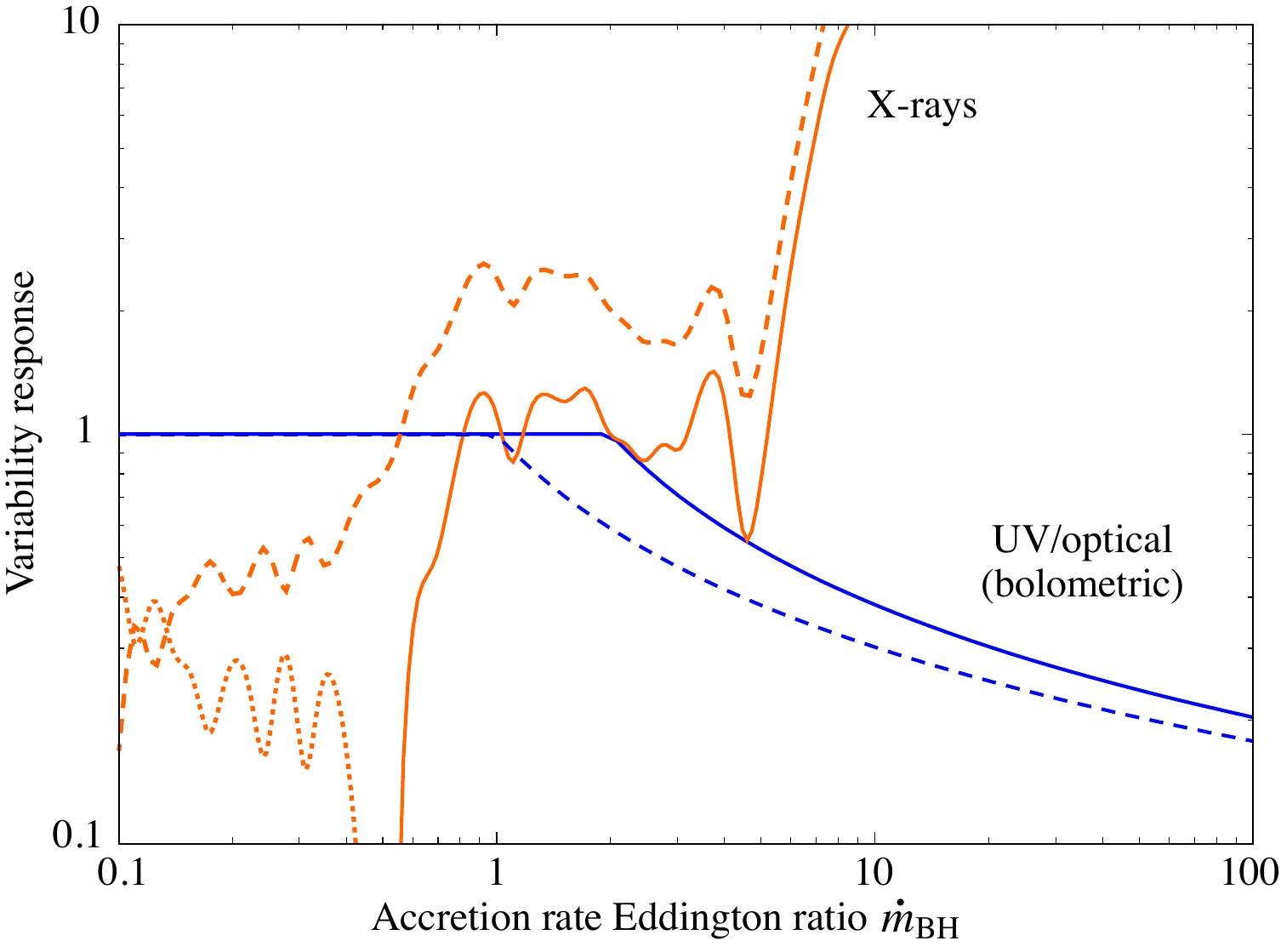}
\end{center}
    \caption{{\it Left}: The normalized bolometric (blue) and X-ray (orange) luminosities as functions of accretion rate, relative to the Eddington value.
    The bolometric luminosity of the disk is presented by \citet{Watarai_2000} (with a 10\% radiative efficiency; our fiducial model).
    The X-ray luminosity is shown for a BH mass of $\mbh =10^7~\msun$ with $y=1$ (solid) and $y=2/3$ (dashed).
    The X-ray Eddington ratio is rescaled by a factor of $10^3$.  
    {\it Right}: Variability response defined as $\mathscr{R} \equiv \frac{{\rm d}\log L}{{\rm d} \log \dot{m}_{\rm BH}}$ for both bolometric and X-ray luminosities.
    In the super-Eddington regime ($\dot{m}_{\rm BH}>1$), the bolometric (UV/optical) luminosity exhibits weak variability due to photon trapping in
    the dense disk, while the X-ray component shows stronger variability at a given accretion-rate fluctuation. 
    The transition accretion rate depends on the BH spin (the blue dashed curve for a 20\% radiative efficiency, corresponding 
    to $a_{\rm BH}\simeq 0.96$).
    Note that the variability response of X-rays is negative $\mathscr{R}<0$, except at the sub-Eddington regime for $y=1$ (denoted by dotted curve).
    This results in an anti-correlation between flux variations in the UV/optical and X-ray bands: as the accretion rate increases, UV/optical luminosity rises
    while X-ray luminosity declines.
    {Alt text: Graphs showing our modeled results.}}
    \label{fig:l_mdot}
\vspace{-3mm}
\end{figure*}

For lower-mass BHs with $\mbh \simeq 10^{6-7}~\msun$, bolometric luminosities reach only 
$L_{\rm bol}\simeq 10^{44-46}~{\rm erg~s}^{-1}$ under similar Eddington ratios.
Thus, substantially higher accretion rates are required to explain such bright populations.
In these cases, a high Eddington ratio increases the optical depth of the coronal region, which softens the X-ray spectrum and 
increases bolometric correction factor.
This trend is consistently observed in this theoretical model for low-luminosity JWST AGNs, most of which have BH masses of $\sim 10^7~\msun$
\citep{Harikane_2023_agn,Maiolino_2024b}.

For local NLSy1 galaxies (red circles; \cite{Liu_2021}), the X-ray bolometric correction factor generally increases with bolometric luminosity. 
Observed NLSy1 galaxies with $L_{\rm bol}/L_{\rm X,2-10~keV}\sim 100$ have bolometric luminosities of $L_{\rm bol} \gtrsim 10^{45}~{\rm erg~s}^{-1}$, 
corresponding to Eddington ratios of $\lambda_{\rm Edd}\gtrsim 0.1$. 
Some fainter NLSy1 galaxies with $L_{\rm bol} \lesssim 10^{45}~{\rm erg~s}^{-1}$ exhibit stronger X-ray emission than predicted by our model, 
as these are low-accretion-rate sources outside the scope of this study (see the discussion at the end of Section~\ref{sec:XSED}).

Our X-ray weak SED model also provides an insight to the development of the cosmic X-ray background (CXB) at high redshifts.
A high abundance of high-redshift AGNs identified through JWST observations indicates that their cosmic BH accretion rate density 
is more than an order of magnitude higher than that estimated from X-ray detected AGNs at similar redshift of $z\gtrsim 4$ 
\citep{Pouliasis_2024,Inayoshi_Ichikawa_2024}.
Given that the contribution of X-ray detected AGNs at $z \sim 3-5$ to the CXB is only a few percent \citep{Ueda_2014},
the CXB would be overproduced if JWST-detected AGNs emitted X-rays as brightly as local AGNs \citep{Padmanabhan_Loeb_2023}.
Our model, suggesting that JWST-selected AGNs are intrinsically  fainter in X-rays, naturally explains 
the discrepancy between the accretion rate densities observed by JWST and Chandra without conflicting with CXB observations.

\vspace{-3mm}
\subsection{Weak variability of AGN disks \& Anti-correlation between UV/optical and X-ray bands}\label{sec:variability}

Another intriguing characteristic of JWST-identified AGNs is their weak (or absent) time variability.
\citet{Kokubo_Harikane_2024} analyzed multi-epoch observations ($\sim $ 2--4 observations per source) of JWST AGNs 
(2 unobscured and 3 LRD populations) and found that they exhibit no detectable flux variation in multi-band photometric data, 
with an upper limit of $\sim 0.1$ mag on variability.
While this study considers only five sources, this result challenges the conventional view that typical quasars often show flux variations
on the order of $0.1-1$ mag in rest-frame optical bands.
Recently, \citet{Zhang_2024} examined the variability significance in a sample of $\sim 300$ LRDs from publicly available photometric data
of various JWST surveys, and found that the LRD population on average does not show strong variability.
While they indeed identified eight strongly variable LRDs with variability amplitudes of $0.24-0.82$ mag,
the fraction remains as small as $\simeq 2.7\%$ out of the parent sample.
This type of weak variabilities have been previously observed in low-redshift AGNs at high accretion rates, such as NLSy1 galaxies,
which typically show very low level of optical variability (e.g., \cite{Klimek_2004,Du_2016,Lu_2019}), though the physical origin remains debated.

In our model, the UV/optical luminosity originates from the accretion disk.
When the disk accretion rate exceeds the Eddington value, a substantial fraction of radiation energy is advected toward the central BH 
before diffusing outward from the disk \citep{Abramowicz_1988,Kato_2008}.
The left panel of Figure~\ref{fig:l_mdot} presents the relationship between bolometric luminosity (or in UV/optical bands) and accretion rate (blue),
using a Slim disk model from \citet{Watarai_2000} with a 10\% radiative efficiency, corresponding to BH spin parameter of $a_{\rm BH}\simeq 0.674$.
At $\dot{m}_{\rm BH} <1$, this model yields an assumed radiative efficiency of $\epsilon_{\rm rad}\simeq 0.1$, 
but for $\dot{m}_{\rm BH} >1$, the dependence on $\dot{m}_{\rm BH}$ weakens due to photon trapping (see also Equation~\ref{eq:Lbol}).
This logarithmic dependence at high accretion rates is crucial for damping luminosity variations ($\Delta L/L$) in response to fluctuations 
in accretion rates ($\Delta \dot{m}_{\rm BH}/\dot{m}_{\rm BH}$).
To quantify this effect, the right panel of Figure~\ref{fig:l_mdot} shows the variability response $\mathscr{R}$, defined as $\mathscr{R} \equiv 
\frac{{\rm d}\log L}{{\rm d} \log \dot{m}_{\rm BH}}$.
The variability response is close to unity at low $\dot{m}_{\rm BH}$ values, and decreases to $\mathscr{R}\lesssim 0.2$ at higher $\dot{m}_{\rm BH}$ values.
The flux variation (in units of AB magnitude) is given by
\begin{align}
\Delta m_{\rm AB}&=2.5\log \left[1+\mathscr{R}(\dot{m}_{\rm BH})\frac{\Delta \dot{m}_{\rm BH}}{\dot{m}_{\rm BH}}\right],\nonumber\\
&\simeq 2.5 \log e \cdot \mathscr{R}(\dot{m}_{\rm BH})\frac{\Delta \dot{m}_{\rm BH}}{\dot{m}_{\rm BH}},
\end{align}
where $|\Delta \dot{m}_{\rm BH}/\dot{m}_{\rm BH}|\ll1$.
Thus, in the sub-Eddington regime ($\mathscr{R}\sim 1$), the flux variation observed for bright quasars ($\Delta m_{\rm AB}\sim 0.1$ mag) can be 
explained by variations in accretion rates with $\Delta \dot{m}_{\rm BH}/\dot{m}_{\rm BH} \simeq 0.1$.
At the super-Eddington regime of $\dot{m}_{\rm BH} >1$, assuming similar amplitudes of accretion rates, the flux variation is reduced as 
$\mathscr{R}$ decreases due to effective photon trapping.
This hypothesis can be quantitatively examined by comparing the variability significance distribution of observed sources with
those predicted by the model (see Figure~15 of \cite{Zhang_2024}).

The characteristic accretion rate above which $\mathscr{R}$ decreases depends on the BH spin and its associated radiative efficiency.
As the BH spin increases to $a_{\rm BH}\simeq 0.96$, the radiative efficiency doubles, thereby reducing the transition accretion rate 
by the same factor (see the blue dashed curve in the right panel of Figure~\ref{fig:l_mdot}).
Consequently, super-Eddington accretion disks at $\dot{m}_{\rm BH}\gtrsim 1$ around rapidly spinning BHs 
show a weaker luminosity response.

Figure~\ref{fig:l_mdot} also shows the Eddington ratio and variability response in the 2-10 keV X-ray band (orange)\footnote{
The X-ray response curves are derived from the super-Eddington coronal model. While the small-scale oscillating features seen in the curves arise from 
numerical differentiation of discrete data points and not physically significant, the overall trend is robust; namely, the X-ray response is weak at low $\dot{m}_{\rm BH}$ 
and becomes stronger as $\dot{m}_{\rm BH}$ increases.}.
Here, we adopt SED models with a BH mass of $\mbh =10^7~\msun$ for $y=1$ (solid) and $y=2/3$ (dashed), as representative cases
(note that the BH mass dependence is much weaker than the dependence on the Compton $y$-parameter).
As seen in Figure~\ref{fig:SED_M}, the X-ray Eddington ratio tends to decrease with the accretion rate because the corona temperature substantially
drops due to higher optical depths. 
Indeed, the variability response becomes larger ($\mathscr{R}\ll -1$) at $\dot{m}_{\rm BH} > 5$, where the optical depth in the corona region reaches 
$\taue\sim 10$ and the electron temperature drops to $kT_{\rm e}\simeq 1$ keV (e.g., a warm corona model; \cite{Kawanaka_Mineshige_2024}).
The quadratic dependence of $kT_{\rm e} \propto \tau_{\rm e}^{-2}$ in the optically-thick regime leads to a significant reduction in temperature,
when keeping the $y$-parameter constant. 
As a result, the luminosity in a specific X-ray band (e.g., rest-frame $2-10$ keV) is dramatically diminished due to variations in accretion rates 
(or equivalently, fluctuations in optical depths).

The UV/optical variability would be arise not only from the intrinsic emission of the accretion disk but also from the reprocessing of variable X-ray irradiation by the corona.
In the super-Eddington regime ($\dot{m}_{\rm BH}\gtrsim 3$), the hard X-ray luminosity contributes at most $\lesssim 1\%$ of the UV/optical luminosity, as shown in Figure~\ref{fig:l_mdot}.
Therefore, even if the hard X-ray luminosity varies significantly, UV/optical variability driven by X-ray heating of the disk would likely remain weaker than the observed upper limit
of variability amplitude for LRDs, $\sim 10\%$ in the rest-frame UV/optical bands \citep{Kokubo_Harikane_2024,Zhang_2024}.

This model robustly predicts an anti-correlation between the flux evolution in the UV/optical and X-ray bands.
As the accretion rate enters the super-Eddington regime, the UV/optical luminosity from the disk increases while the X-ray luminosity decreases.
Observing this anti-correlation may be challenging without long-term monitoring that extends beyond the typical decorrelation timescale 
(e.g., \cite{Kozlowski_2017}).
Recently, \citet{R.Li_2024a} conducted a three-year monitoring campaign for the changing-look AGN 1ES 1927+654, covering X-ray to UV/optical wavelengths,
and reported an outburst triggered by a stellar tidal disruption event (TDE) at super-Eddington accretion rates.
As the accretion rate decreased following $\dot{M}\propto t^{-1.53}$, the disk luminosity declined at slower rate (possibly due to photon trapping).
Meanwhile, the hard X-ray component disappeared around $\sim 200$ days after the outburst, possibly due to
the destruction of the corona caused by shocks between debris from a tidally disrupted star
and the accretion flow (see the discussion in \cite{Ricci_2020}).
Following this temporary disappearance, the hard X-ray luminosity rose sharply
and then continued to increase until the Eddington ratio dropped below unity.
After entering the sub-Eddington regime, the X-ray luminosity began to decline.
The overall evolutionary trend of the broadband SED for this changing-look AGN is well explained by our SED model
(see also their SED model and physical interpretation; \cite{R.Li_2024b}).

\vspace{-3mm}
\subsection{Uniqueness of the high-$z$ universe}\label{sec:unique}

In this work, we propose a possible solution to the weakness of X-rays and UV/optical variability for JWST-identified AGNs 
by suggesting that many of these AGNs are accreting at super-Eddington rates.
This naturally raises a question: why are super-Eddington accretors more common at higher redshifts? 
Below, we address this question with a simple analytical model.

To evaluate the Eddington ratio of the BH accretion rate, we begin by estimating the mass growth rate of the host dark matter (DM) halo.
Cosmological $N$-body simulations have shown that the mass assembly of DM halos is well described by a functional form,
$M_{\rm h}\propto e^{-k_{\rm h}z}$ \citep{Wechsler_2002,Neistein_Dekel_2008}, where the mean value of $\langle k_{\rm h}\rangle$ is calibrated to $\simeq 0.7$ \citep{Fakhouri_2010}.
The redshift dependence leads to a growth rate for the halo mass that scales as ${\rm d}(\ln M_{\rm h})/{\rm d}t \propto (1+z)^{5/2}$,
which is also derived from the extended Press-Schechter formalism and a fit to merger trees from cosmological simulations
\citep{Fakhouri_2010,Dekel_2013}.
Assuming a constant star formation efficiency (conversion of gas into stars), the stellar-mass growth rate can be expressed as
\begin{align}
\dot{M}_{\star}
\simeq H_0\sqrt{\Omega_{\rm m}}(1+z)^{5/2}k_{\rm h}M_{\star},
\end{align}
where $z\gg (\Omega_\Lambda/\Omega_{\rm m})^{1/3}-1\simeq 0.3$ is considered.
Next, we assume that the BH growth is linked to the stellar mass growth rate, implying $\dot{M}_{\rm BH}= (dM_{\rm BH}/dM_\star)\dot{M}_\star$, 
where the redshift dependence is taken from $M_\star(z)$.
As a result, the Eddington ratio of the BH growth rate is expressed as 
\begin{align}
\dot{m}_{\rm BH}\simeq 0.55~ k_{\rm h}\cdot 
\frac{{\rm d}\ln M_{\rm BH}}{{\rm d}\ln M_{\star}}\left(\frac{1+z}{10}\right)^{5/2},
\label{eq:halo}
\end{align}
where the derivative term $\mathscr{D}=\frac{{\rm d}\ln M_{\rm BH}}{{\rm d}\ln M_{\star}}$ describes the evolution of 
the BH-to-stellar mass ratio $M_{\rm BH}/M_\star$: 
for $\mathscr{D}>1$, the BH growth increases the $M_{\rm BH}/M_\star$ ratio;
for $\mathscr{D}<1$, the mass ratio declines over time; and 
for $\mathscr{D}=1$, the $M_{\rm BH}/M_\star$ ratio remains constant through the mass evolution.

This equation~(\ref{eq:halo}) provides two key insights.
First, as the redshift increases, the Eddington ratio tends to rise following the $(1+z)^{5/2}$ dependence.
Additionally, the value of $k_{\rm h}$ has a broader distribution with a long tail of $k_{\rm h}\gtrsim \langle k_{\rm h}\rangle$ 
at higher redshifts \citep{Fakhouri_2010}. 
Thus, a significant fraction of DM halos at high redshifts have a value larger than $\langle k_{\rm h}\rangle \simeq 0.7$,
which enhances the Eddington ratio for the BH accretion rate.
These facts suggest that the BH growth rate, which is linked to the gas supply from the host DM halos, approaches or exceeds 
the Eddington value preferentially at earlier cosmic epochs.
Second, the Eddington ratio increases when the BH-to-stellar mass ratio increases ($\mathscr{D} > 1$).
This is likely the case for JWST-identified AGNs that show high $M_{\rm BH}/M_\star$ ratios at their observed epochs of $z\sim 4-7$
(e.g., \cite{Kocevski_2023,Harikane_2023_agn,Maiolino_2023_JADES,Chen_2024}), 
especially when their seed BHs started with lower $M_{\rm BH}/M_\star$ ratios in the seeding epochs at $z>10$.

Considering these insights, we propose a scenario that a majority of high-redshift BHs (particularly lower mass ones) that experience 
super-Eddington accretion tend to become overmassive relative to the local BH-to-stellar mass ratio (e.g., \cite{Inayoshi_2022a,Hu_2022b,Scoggins_2023}).
Furthermore, this scenario provides a consistent explanation for the observed weakness of X-rays and UV/optical variability 
in these high-redshift AGNs, as super-Eddington accreting BHs diminish hard X-rays and damp variability in the same manner 
as their lower-redshift counterparts.

%%%%%%%%%%
%	Section 5     %
%%%%%%%%%%
\vspace{-3mm}
\section{Summary}\label{sec:summay}

The JWST observations have led to the exploration of broad-line AGNs in the early universe at redshifts $z\gtrsim 4-7$. 
While these sources show clear radiative and morphological signatures of AGNs in rest-frame optical bands, complementary evidence 
of AGN activity -- such as X-ray emission and UV/optical variability -- remains rarely detected (e.g., \cite{Maiolino_2024b,Yue_2024,Kokubo_Harikane_2024}). 
The weakness of X-rays and variability in these AGNs challenges the conventional AGN paradigm, suggesting 
that the accretion and radiative mechanisms around the central BH may differ from those of low-redshift counterparts
and high-redshift more luminous populations.

In this paper, we investigate the radiation spectra of super-Eddington accretion disks surrounded by dense coronae. 
In the vicinity of these disks, radiation-driven outflows transport mass to the polar regions, leading to the formation of 
moderately optically-thick, warm coronae through effective inverse Comptonization.
This mechanism results in softer X-ray spectra and larger bolometric correction factors for X-rays compared to typical AGNs, 
while aligning with those of JWST AGNs \citep{Maiolino_2024b,Yue_2024} and low-redshift super-Eddington 
accreting AGNs \citep{Laurenti_2022}.
Our broadband SED model explains the X-ray faintness (or non-detection) of JWST-identified AGNs powered by 
lower-mass BHs with $M_{\rm BH}\lesssim 10^{7-8}~\msun$.

In this scenario, UV/optical variability is suppressed due to photon trapping within super-Eddington disks, while X-ray emissions 
remain weak but highly variable.
Observing this anti-correlation may require long-term monitoring of AGNs, though particular transient phenomena such as
stellar TDEs could serve as a critical test for this model.
Wide-field surveys, such as JWST COSMOS-Web \citep{Casey_2023} and upcoming surveys, NEXUS \citep{Shen_2024}, 
the High Latitude Time Domain Survey with the Roman Space Telescope \citep{Rose_2021}, and 
the Legacy Survey of Space and Time (LSST) by the Vera C. Rubin Observatory \citep{Bianco_2022},
could offer ideal platforms to observe these events especially high-redshift TDEs \citep{Inayoshi_2024}.

These unique characteristics are expected to appear preferentially in higher redshift AGNs that host accreting, lower-mass 
BHs with $\lesssim 10^{7-8}~\msun$.
Particularly, if seed BHs began growing from lower $M_{\rm BH}/M_\star$ ratios in the seeding epochs at $z>10-20$,
super-Eddington accretion would be necessary to achieve the BH-to-stellar mass ratio to $\sim 0.1$, as observed for 
JWST-identified AGNs at $z\sim 4-7$.
Under the independent requirement of rapid BH growth, the observed weakness of X-ray and UV/optical variability 
in these high-redshift AGNs are consistently explained.

\section*{Acknowledgments}
We greatly thank Changhao Chen, Zolt\'an Haiman, Yuichi Harikane, Luis C. Ho, Kunihito Ioka, Mitsuru Kokubo, Ruancun Li, Piero Madau, Roberto Maiolino, 
Tatsuya Matsumoto, Kohta Murase, Raffaella Schneider, Jinyi Shangguan, Rosa Valiante, Luca Zappacosta, and Zijian Zhang 
for constructive discussions.
K.~I. acknowledges support from the National Natural Science Foundation of China (12073003, 12003003, 11721303, 11991052, 11950410493), 
and the China Manned Space Project (CMS-CSST-2021-A04 and CMS-CSST-2021-A06). 
S.~S.~K. acknowledges the support by KAKENHI No.~22K14028, 21H04487, and 23H04899, and  Tohoku Initiative for Fostering Global Researchers for Interdisciplinary Sciences (TI-FRIS) of MEXT's Strategic Professional Development Program for Young Researchers.
H.~N. acknowledges the support  by KAKENHI No.~23K20239 and 24K00672.
We thank the Yukawa Institute for Theoretical Physics at Kyoto University, where this study was initiated during the YITP-W-24-22 on ``Exploring Extreme Transients".
This research was also supported in part by grant NSF PHY-2309135 to the Kavli Institute for Theoretical Physics.

\appendix
\section{Energy conservation in disk + corona systems}\label{sec:appendix1}

Our corona model is based on a simplified energy conservation law between the disk and corona \citep{Kawanaka_2021,Kawanaka_Mineshige_2024}.
A recent study by \citet{Madau_Haardt_2024} addresses the X-ray weakness of JWST-identified AGNs by considering the effect
of super-Eddington accretion disks, especially focusing on warm coronae cooled by soft photons in a funnel-like, reflective geometry 
above a geometrically thick disk with a scale-height of $H/R\simeq 1$.

The energy conservation equation adopted in \citet{Madau_Haardt_2024} differs from the Equation~(\ref{eq:energy_soft}) 
by adding incoming radiation from other parts of the funnel wall 
\begin{equation}
    (1-f_w) Q^+_{\rm tot} + yF_{\rm soft}/2 +\mathscr{G}e^{-2\taue}Q^+_{\rm tot}= F_{\rm soft},
    \label{eq:energy_MH24}
\end{equation}
where the factor $e^{-2\taue}$ accounts for the fraction of the radiation flux that, after being scattered in the corona region, 
reaches the funnel wall from the opposite side, and $\mathscr{G}$ represents a geometrical factor related to the steepness of the funnel structure.
As the opening half-angle $\Theta$ of the funnel, defined by $\Theta\equiv {\rm arccot}(H/R)_{\rm max}$, becomes smaller due to a
thicker disk structure (i.e., higher Eddington ratios), the $\mathscr{G}$ factor increases and reaches $\mathscr{G}\sim 2-3$ for 
$\dot{m}_{\rm BH}\sim 10$ \citep{Madau_Haardt_2024}.  
Using Equation~(\ref{eq:energy_MH24}), the relationship between $f_w$ and $y$ is modified from Equation~(\ref{eq:fwy}) as
\begin{equation}
    y=\frac{2f_w}{1+\mathscr{G}e^{-2\taue}-f_w}.
    \label{eq:fwy_MH24}
\end{equation}
Thus, for a fixed $f_w$ value, the corresponding Compton $y$-parameter is reduced.
By equating Equation~(\ref{eq:fwy_MH24}) with $y=(4\Te+16\Te^2)\taue(\taue+1)$, the electron temperature is lower
than the case with $\mathscr{G}=0$.

This non-local heating effect owing to a thick disk structure plays an important role in lowering the corona temperature,
when the flux suppression factor $e^{-2\taue}$ is moderate for optically-thin coronae ($\taue<1$).
In contrast, a larger value of $\mathscr{G}\sim 2-3$ is achieved at higher accretion rates ($\dot{m}_{\rm BH}\sim 10$), 
corresponding to $\taue\sim 20$ when radiation-driven outflows from the super-Eddington accretion disk are considered
(see Equation~\ref{eq:tau1}).
In this scenario, the value of $\mathscr{G}e^{-2\taue}$ becomes negligibly small.

\section{Comptonization model}\label{sec:LG}

%%%%%%%%
%   Figure 5    %
%%%%%%%%
\begin{figure}
\begin{center}
\includegraphics[width=83mm]{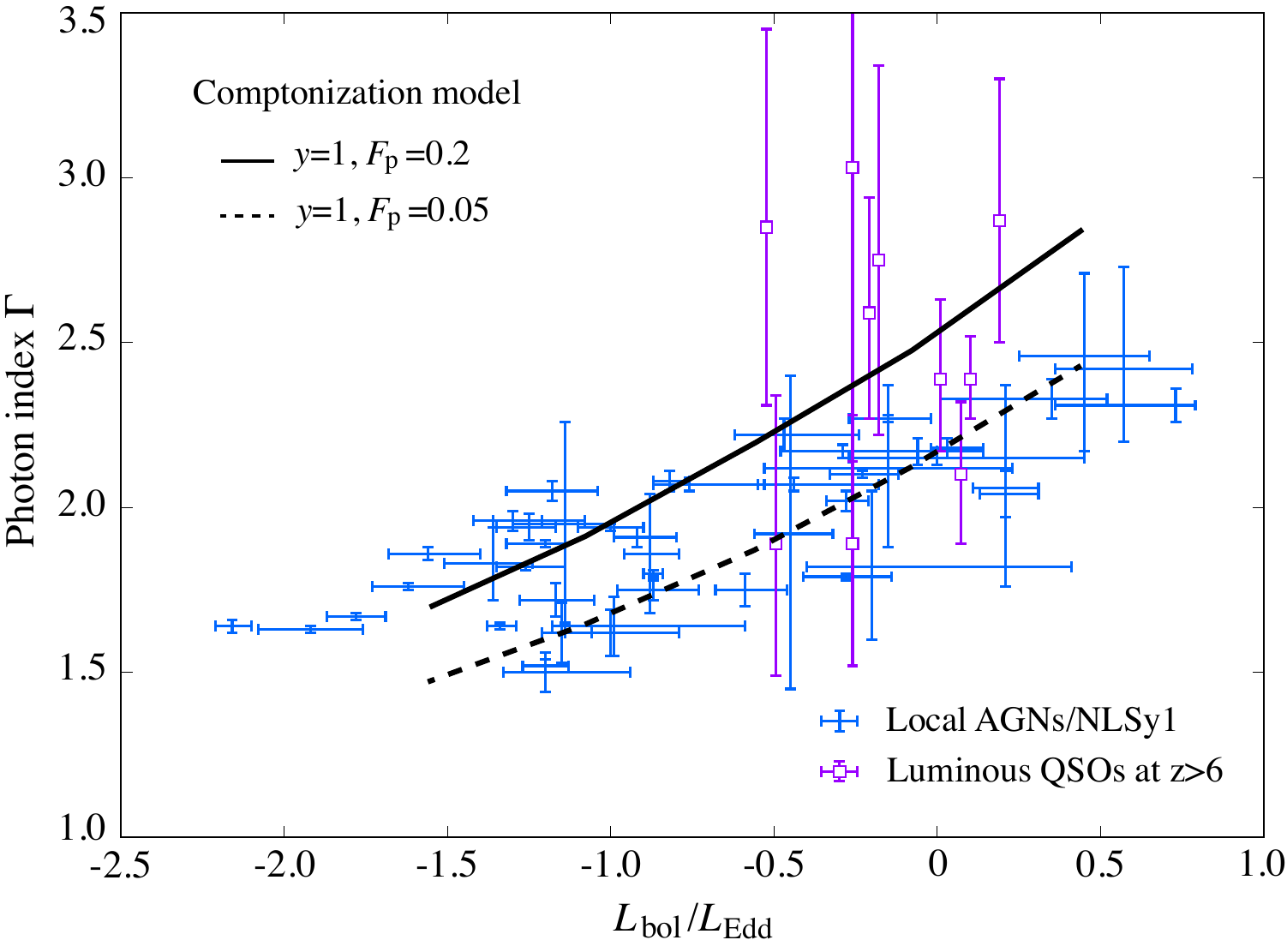}
\end{center}
    \caption{The $L_{\rm bol}/L_{\rm Edd}-\Gamma$ relationship based on the Comptonization model adopted in this work.
    Here, the Comptonization $y$-parameter and mass-loading factor are given as $y=1$ with $F_{\rm p}=0.2$ (black solid) and $F_{\rm p}=0.05$ (black dashed). 
    Observation data for various types of AGNs from literatures are overlaid for comparison: local AGNs including NLSy1 galaxies (blue symbols, \cite{Liu_2021}) 
    and luminous quasars at $z>6$ (purple symbols, \cite{Zappacosta_2023})}
    \label{fig:LG}
\vspace{-3mm}
\end{figure}

In our model framework, the relationship between the Eddington ratio ($L_{\rm bol}/L_{\rm Edd}$) and photon index ($\Gamma$) is calculated using 
Equations~(\ref{eq:Gamma1}) and (\ref{eq:beta}). 
Figure~\ref{fig:LG} presents this relationship for a fixed Compton $y$-parameter of $y=1$ with two different mass-loading factors: 
$F_{\rm p}=0.2$ (black solid) and $F_{\rm p}=0.05$ (black dashed).
We overlay observational data for local AGNs, including NLSy1 galaxies (blue symbols; \cite{Liu_2021}), 
and luminous quasars at $z>6$ (purple symbols; \cite{Zappacosta_2023}).
For consistency with \citet{Liu_2021}, we apply a bolometric correction factor of 2.75 to convert modeled UV luminosities to bolometric luminosities.
Our fiducial model with ($F_{\rm p}=0.2$) matches well with the sub-Eddington sources reported in \citet{Liu_2021}, while also predicting a steeper photon 
index in the super-Eddington regime, consistent with the high-redshift quasars of \citet{Zappacosta_2023}.
The lower mass-loading model ($F_{\rm p}=0.05$) better reproduces the observed trend for the super-Eddington AGN samples compiled in \citet{Liu_2021}.

\end{document}